# Gravitational lensing and shadow of charged black holes in the low-energy limit of string theory

Younes Younesizadeh[1,a], Feyzollah Younesizadeh[2], Mohammad M. Qaemmaqami[3]

[1] Department of Physics, Isfahan University of Technology, Isfahan, Iran
[2] Physics Department, Amirkabir University of Technology, Tehran, Iran
[3] School of Particles and Accelerators, Institute for Research in Fundamental Sciences (IPM), P.O. Box 19395-5531, Tehran, Iran

**Abstract** In this work, we investigate the shadow cast and strong field gravitational lensing of a new class of black hole solutions in dilaton gravity where dilaton field is coupled with nonlinear Maxwell invariant [Younesizadeh et al. in Int J Mod Phys A 34(35):1950239]. The space-time is a stationary axisymmetric geometry. The key part in our investigations is finding the effect of dilaton parameter $N$ on the size of shadows and the energy emission rate. As the $N$ parameter increases, the size of black hole shadow increases. Also, the energy emission rate increases with increase in the dilaton parameter $N$. By supposing the gravitational field of the supermassive object at the heart of Milky Way galaxy described by this metric, we estimated the numerical values of the observables for gravitational lensing in the strong field limit.

## 1 Introduction

One of the most interesting and bizarre objects in the universe is black holes. After a lot of different astrophysical research in the past years, astronomers believe in the existence of supermassive black holes in heart of the most galaxies. At least we can address to the two of them as the Milky Way and neighboring elliptical M87 galaxies [1,2]. The breakthrough in this regard is the Event Horizon Telescope (EHT) Collaboration which they announced the first image of a supermassive black hole located at the center of the giant elliptical galaxy M87 [3]. For this purpose, they used a technique called Very Long Baseline Interferometry (VLBI). As we know, the gravitational attraction around a black hole is so intense that the nearby objects will fall into it when they reach a critical radius named $r_c$. These astounding phenomena are known as the gravitational lensing. When the nearby objects are photons emitted by an illuminated source located behind the black hole, then it brings a shadow which can be seen by an observer sitting at infinity. The shadow concept has been studied first by Bardeen [4]. Usually, the shadow of spherically symmetric black holes is circular [5], but for the spinning black holes this is not true and the shadow is actually deformed [6,7].

[a] e-mail: younesizadeh88@gmail.com (corresponding author)

Scalar fields play an important role in theoretical physics. A crucial one is the dilaton field which is coupled to the Einstein gravity in the low energy limit of string theory. In past decades, this scalar field has been seen in different theories such as dimensional reduction, various models of supergravity, low-energy limit of string theory [8,9]. Also, we note that the effect of dilaton field can change the structures of the solutions, the causal structure of space-time geometry, the asymptotic behavior of the space-time, and the thermodynamical properties of charged black hole (BH) solutions [10,11].

The power-law Maxwell electrodynamics is a famous of nonlinear electrodynamics models. As we know, the electromagnetic part of its Lagrangian density is $\mathcal{F} = (F_{\mu\nu}F^{\mu\nu})^p$.[1] Lagrangian is invariant under this conformal transformation $g_{\mu\nu} \to \Omega^2 g_{\mu\nu}$ and $A_{\mu\nu} \to A_{\mu\nu}$. These days, the power of Maxwell invariant (PMI) has been considered as an important area to study in the background of geometrical physics [12–14] and more obviously in the framework of Einstein–Maxwell dilaton gravity [15,16].

The paper is organized as the following order. In Sect. 2, we first introduce the Einstein-Generalized Maxwell Lagrangian with a dilatonic field coupling. Then, from varying this action with respect to the metric $g_{\mu\nu}$, dilaton field $\Phi$, and the gauge field $A_\mu$, we have introduced the equations of motion. Also, we have solved these equations in order to obtain the black holes solution. In the end of this section, we have also showed that the dilaton potential can be written as the generalized three Liouville-type potentials. In Sect. 3, we have obtained the black hole shadow cast. In Sect. 4, we discuss the energy emission rate in the high energy case from the dilatonic black hole solution. In Sect. 5, we have introduced the gravitational lensing and calculate the observables corresponding to the supermassive Galactic black hole at the center of our galaxy. Final section is devoted to summary and closing remarks.

## 2 Field equations and the BH solutions

In this section, we introduce the Einstein-Generalized Maxwell gravity which is coupled to a dilaton field. The general form of this action is as [17,18]:

$$S = \frac{1}{16\pi} \int d^4x \sqrt{-g} \left( \mathcal{R} - 2g^{\mu\nu}\nabla_\mu \Phi \nabla_\nu \Phi - V(\Phi) + \mathcal{L}(\mathcal{F}, \Phi) \right) \qquad (1)$$

where $\mathcal{R}$ is the Ricci scalar, $V(\Phi)$ is the potential of dilaton field. The last term is the coupling between electrodynamics and scalar field. This term is as $\mathcal{L}(\mathcal{F}, \Phi) = \left(-\mathcal{F}e^{-2\alpha\Phi}\right)^p$ where $\mathcal{F} = F_{\mu\nu}F^{\mu\nu}$ (In electromagnetic context, $F_{\mu\nu}$ is defined as $F_{\mu\nu} = \partial_\mu A_\nu - \partial_\nu A_\mu$), $\alpha$ is the coupling constant and power p is the nonlinearity parameter. In the case of $p = 1$, this action brings the Einstein–Maxwell-dilaton gravity action. Varying this action with respect to the metric $g_{\mu\nu}$, dilaton field $\Phi$ and the gauge field $A_\mu$ yields:

$$\mathcal{R}_{\mu\nu} = 2\partial_\mu \Phi \partial_\nu \Phi + \frac{1}{2}g_{\mu\nu}V(\Phi) + \left[\left(p - \frac{1}{2}\right)g_{\mu\nu} + \frac{p}{\mathcal{F}}F_{\mu\gamma}F_\nu^\gamma\right]\mathcal{L}(\mathcal{F}, \Phi) \qquad (2)$$

$$\nabla_\mu(\sqrt{-g}\mathcal{L}_\mathcal{F}(\mathcal{F}, \Phi)F^{\mu\nu}) = 0 \qquad (3)$$

$$\partial_\mu \partial^\mu \Phi = \frac{1}{4}\frac{\partial V(\Phi)}{\partial \Phi} + \frac{\alpha p}{2}\mathcal{L}(\mathcal{F}, \Phi) \qquad (4)$$

And $F_{\mu\gamma}F_\nu^\gamma \equiv g^{\beta\gamma}F_{\nu\beta}F_{\mu\gamma}$, $\mathcal{L}_\mathcal{F}(\mathcal{F}, \Phi) \equiv \frac{\partial}{\partial \mathcal{F}}\mathcal{L}(\mathcal{F}, \Phi)$.

---

[1] $p$ is the nonlinearity parameter.

Now, we consider the following ansatz metric

$$ds^2 = -X(r)dt^2 + \frac{dr^2}{X(r)} + 2aH(r)\sin^2\theta dt d\phi + f(r)^2 d\Omega_2^2 \tag{5}$$

where $d\Omega_2^2$ is the line element of a two-dimensional hypersurface. Also, $X(r)$ $H(r)$ and $f(r)$ are three unknown functions of $r$ that should be found. For small rotation case, we are able to solve Eqs. (2)–(4) to first order in angular momentum parameter $a$. By solving Einstein field equations, the only term which is added to non-rotating case is $g_{t\phi}$ that contains the first order of $a$.

Here, we consider this gauge potential:

$$A_\mu = h(r)\bigl(\delta_\mu^t - a\sin^2\theta \delta_\mu^\phi\bigr) \tag{6}$$

Integration of Maxwell equation [Eq. (3)] gives:

$$F_{tr} = \frac{qe^{\frac{2\alpha p \Phi(r)}{2p-1}}}{f(r)^{\frac{2}{2p-1}}} \tag{7}$$

where $q$ is an integration constant in this relation which is related to the electric charge.

By using the above ansatz [Eq. (5)] in Eq. (2), one can find the following field equations. These field equations are for $tt$, $rr$, $\theta\theta$ and $t\phi$ components, respectively.

$$2f'(r)X'(r) + f(r)X''(r) = f(r)V(\Phi) + (3p-1)f(r)\mathcal{L}(\mathcal{F}, \Phi) \tag{8}$$

$$2f'(r)X'(r) + f(r)X''(r) + 4f''(r)X(r) = 4f(r)X(r)\Phi'^2 + f(r)V(\Phi) \\ + (3p-1)f(r)\mathcal{L}(\mathcal{F}, \Phi) \tag{9}$$

$$\bigl[X(r)(f^2(r))'\bigr]' - 2 = f^2(r)V(\Phi) + (2p-1)f^2(r)\mathcal{L}(\mathcal{F}, \Phi) \tag{10}$$

$$f^2(r)X(r)H''(r) + 2H(r)\bigl(f(r)f'(r)X'(r) - 1\bigr) = f^2(r)H(r)V(\Phi) \\ + \bigl((2p-1)H(r) + pX(r)\bigr)f^2(r)\mathcal{L}(\mathcal{F}, \Phi) \tag{11}$$

From Eqs. (8) and (9), we have

$$\frac{f''(r)}{f(r)} = \Phi'(r)^2 \tag{12}$$

We further assume this ansatz as well:

$$f(r) = \beta r^N \tag{13}$$

where $\beta$ and $N$ are just two constants. By putting this ansatz in Eq. (12), $\Phi(r)$ is:

$$\Phi(r) = \pm\sqrt{N(N-1)}\ln(r) + \Phi_0 \tag{14}$$

$\Phi_0$ is an integration constant. Without the loss of generality, we set $\Phi_0 = 0$. In this relation, $N$ parameter is related to the dilaton field and if we set $N = 1$, dilaton field will be vanished ($N \geq 1$).

From Eqs. (6, 7, 13, 14), the non-vanishing components of the electromagnetic field tensor are:

$$F_{tr} = -F_{rt} = \begin{cases} \dfrac{q}{\beta^{\frac{2}{2p-1}} r^{\frac{2}{2p-1}}} & \text{for } N = 1 \\ \dfrac{q}{\beta^{\frac{2}{2p-1}}} r^{\frac{2N[2p(N-1)-1]}{2p-1}} & \text{otherwise} \end{cases} \quad (15)$$

$$F_{\phi r} = -a \sin\theta F_{tr}, \quad F_{\theta\phi} = -ah(r) \sin 2\theta \quad (16)$$

For $h(r)$ function in the gauge potential Eq. (6), we have:

$$h(r) = \begin{cases} \dfrac{(2p-1)q}{\beta^{\frac{2}{2p-1}}(2p-3)} r^{\frac{2p-3}{2p-1}} & \text{for } N = 1 \\ \dfrac{(2p-1)q}{\beta^{\frac{2}{2p-1}}[2p(2N^2-2N+1)-3]} r^{\frac{2p(2N^2-2N+1)-3}{2p-1}} & \text{otherwise} \end{cases} \quad (17)$$

We have assumed $\alpha = 2\sqrt{N(N-1)}$.

From the above equations, one can obtain [19]:

$$X(r) = \frac{r^{2-2N}}{\beta^2(2N-1)} - mr^{1-2N} - \frac{\Lambda}{N(4N-1)} r^{2N} + \Upsilon(r) \quad (18)$$

$$H(r) = -mr^{1-2N} - \frac{\Lambda}{N(4N-1)} r^{2N} + \Upsilon(r) \quad (19)$$

where

$$\Upsilon(r) = \begin{cases} -\dfrac{2^p(2p-1)^2 q^{2p} r^{-\frac{2}{2p-1}}}{4\beta^{\frac{4p}{2p-1}}(2p-3)} & \text{for } N = 1 \\ \dfrac{2^p(2p-1)^2 pq^{2p} r^{2\frac{(2p(N-1)^2-1)}{2p-1}}}{2\beta^{\frac{4p}{2p-1}}(2pN^2-6pN+N+2p-1)(4pN^2-4pN-2N+2p-1)} & \text{otherwise} \end{cases} \quad (20)$$

Without the loss of generality, we can set $\beta = 1$ in the above equations.

One property of these solutions is that if we set $N = 1$, the slowly rotating Kerr metric (Lense–Thirring metric) recovers. Also, by setting $a = 0$ and $p = 1$ the Reissner–Nordstrom metric recovers, too [20].

If dilaton potential considers as a three Liouville-type potentials as follows:

$$V(\Phi) = 2\Lambda_1 e^{2\zeta_1 \Phi} + 2\Lambda_2 e^{2\zeta_2 \Phi} + 2\Lambda e^{2\zeta_3 \Phi} \quad (21)$$

Then, we have

$$\Lambda_1 = -\frac{N-1}{\beta^2(2N-1)}, \quad \Lambda_2 = -\frac{(2p-1)2^{p-1} q^{2p}(2pN^2 - 7pN + N + 2p - 1)}{\beta^{\frac{4p}{2p-1}}[(2N^2 - 6N + 2)p + N - 1]} \quad (22)$$

$$\zeta_1 = -\sqrt{\frac{N}{N-1}}, \quad \zeta_2 = \frac{2p(N-2)}{2p-1}\sqrt{\frac{N}{N-1}}, \quad \zeta_3 = \sqrt{\frac{N-1}{N}} \quad (23)$$

In Eq. (21), $\Lambda$ is a free parameter which plays the role of the cosmological constant.

## 3 Shadow of 4D black holes

The spherically symmetric 4D black hole in the previous section is as follows when we set $a = 0$.

$$ds^2 = -X(r)dt^2 + \frac{dr^2}{X(r)} + f(r)^2\left(d\theta^2 + \sin^2\theta d\phi^2\right) \tag{24}$$

where $X(r)$ and $f(r)$ are given by Eqs. (18) and (13).

Equations of motion for the evolution of the test particle around the black hole can be introduced through the following relations:

$$\begin{aligned}\frac{dt}{d\sigma} &= \frac{E}{X(r)} \\ \frac{d\phi}{d\sigma} &= \frac{L_\phi}{r^2 \sin^2\theta}\end{aligned} \tag{25}$$

In order to find the null geodesic equations, one can use the Hamilton–Jacobi equation

$$\frac{dS}{d\sigma} = -\frac{1}{2}g^{\mu\nu}\frac{\partial S}{\partial x^\mu}\frac{\partial S}{\partial x^\nu} \tag{26}$$

where $\sigma$ is the affine parameter along the geodesics, and $S$ is the Jacobi action. To solve the Hamilton–Jacobi equation, we propose the following ansatz

$$S = \frac{1}{2}m^2\sigma - Et + L_\phi\phi + S_r(r) + S_\theta(\theta) \tag{27}$$

where $m$ is the test particle's mass, $E$ is the energy and $L$ is the angular momentum. $S_r(r)$, $S_\theta(\theta)$ are the function of $r$ and $\theta$.

By substituting the Jacobi action Eq. (27) into the Hamilton–Jacobi equation Eq. (26), we obtain

$$-\frac{E^2}{X(r)} + X(r)\left(\frac{\partial S_r(r)}{\partial r}\right) + \frac{1}{r^{2N}}\left(\frac{\partial S_\theta(\theta)}{\partial \theta}\right) + \frac{L_\phi^2}{r^{2N}\sin^2\theta} = 0 \tag{28}$$

Eq. (28) can be recast into the following form

$$S_r(r) = \int^r \frac{\sqrt{\mathcal{R}(r)}}{r^2 X(r)}dr \tag{29}$$

$$S_\theta(\theta) = \int^\theta \sqrt{\Theta(\theta)}d\theta \tag{30}$$

where the two functions $\mathcal{R}(r)$ and $\Theta(\theta)$ are given by:

$$\mathcal{R}(r) = E^2 r^4 - r^{4-2N} X(r)\left(\mathcal{K} + L_\phi^2\right) \tag{31}$$

$$\Theta(\theta) = \mathcal{K} - \left(L_\phi^2 \cot^2\theta\right) \tag{32}$$

$\mathcal{K}$ is a constant of separation called Carter constant. In Eq. (31), the presence of dilaton field has been changed the form this equation. When we do not have this field $N = 1$, the normal equation recovers [21].

To determine the black hole shadow as seen by the distant observer, we introduce the two celestial coordinates $(x, y)$ [22],

$$x = -\lim_{r_0 \to \infty} r_0 \frac{p^{\hat{\phi}} + p^{\hat{\psi}}}{p^{\hat{t}}}$$

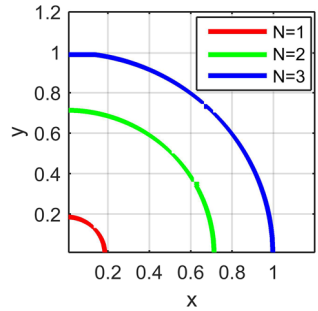

**Fig. 1** Shadow region of the 4D black holes for different values of $m = 100, \Lambda = -1, q = 0$

$$y = \lim_{r_0 \to \infty} r_0 \frac{p^{\hat{\theta}}}{p^{\hat{t}}} \tag{33}$$

The contravariant components of the momenta $p^{\hat{i}}$ in new coordinate basis can be easily computed using the orthonormal basis vectors.

$$p^{\hat{t}} = \frac{E}{X(r)}, \quad p^{\hat{\phi}} = \frac{L_\phi}{r \sin\theta},$$
$$p^{\hat{r}} = \pm\sqrt{X(r)\mathcal{R}(r)}, \quad p^{\hat{\theta}} = \pm\frac{\sqrt{\Theta(\theta)}}{r}, \tag{34}$$

Plunging these equations in Eq. (33) and taking the limit $r_0 \to \infty$, we get

$$x = -\xi_\phi \csc\theta$$
$$y = \pm\sqrt{\eta - \xi_\phi^2 \cot^2\theta} \tag{35}$$

The boundary of shadow of the 4D black holes can be defined through the conditions (Fig. 1)

$$\mathcal{R}(r)\big|_{r=r_p} = \frac{\partial \mathcal{R}(r)}{\partial r}\big|_{r=r_p} = 0 \tag{36}$$

$$\frac{\mathcal{K}}{E^2} + \left(\frac{L_\phi}{E}\right)^2 = \frac{r_p^2}{X(r_p)} \tag{37}$$

$$r_p X'(r_p) - 2N X(r_p) = 0 \tag{38}$$

Where $\frac{\mathcal{K}}{E^2} = \eta$, $\frac{L_\phi}{E} = \xi_\phi$. Also, $\xi_\phi^2 \equiv \xi^2$

From Eq. (35), we obtain

$$x^2 + y^2 = R_s^2 = \eta + \xi_\phi^2 = \frac{r_p^2}{X(r_p)} \tag{39}$$

It can be seen that the shadow size increases with an increase in the dilaton parameter $N$. In this figure (Fig. 2), the red color represents the Reissner–Nordstrom case ($N = 1$) and BHs with $N \geq 2$ have a bigger shadow radius than the RN case.

In this figure (Fig. 3), there is a unique q value that the shadow radius hits its minimum.

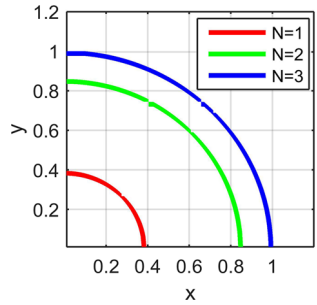

**Fig. 2** Shadow region of the 4D black holes for different values of $m = 100, \Lambda = -1, q = 1, p = 1$

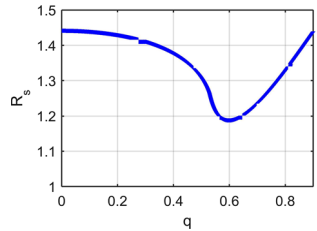

**Fig. 3** Shadow radius of the 4D black holes versus $q$. for different values of $m = 1, \Lambda = -1$, $N = 1, p = 1$

## 4 Energy emission rate

In this section, we will discuss the energy emission rate in the high energy case from the 4D black hole solution in Einstein–Maxwell-dilaton gravity. The expression of energy emission rate reads [23]

$$\frac{d^2 E(\omega)}{d\omega dt} = \frac{2\pi^2 \sigma_{\lim}}{e^{\omega/T_H} - 1} \omega^3 \tag{40}$$

where $E(\omega)$ is the energy, $\omega$ frequency and $T_H$ is the Hawking temperature of the black hole.

The Hawking temperature is:

$$T_H = \frac{(1-N)r_+^{1-2N}}{2\beta^2(2N-1)\pi} - \frac{(2N-1)m}{4\pi r_+^{2N}} - \frac{N\Lambda r_+^{2N-1}}{2N(4N-1)\pi} r^{2N} + \frac{\Upsilon'(r_+)}{4\pi} \tag{41}$$

where $r_+$ is the event horizon radius and $\Upsilon(r)$ is Eq. (20).

The limiting constant value can be expressed in $d$ dimensions as [24].

$$\sigma_{\lim} = \frac{\pi^{\frac{d-2}{2}} R_s^{d-2}}{\Gamma\left(\frac{d}{2}\right)} \tag{42}$$

where $R_s$ is the radius of the black hole shadow. In $d = 4$ dimensions, $\sigma_{\lim}$ can be approximately expressed as:

$$\sigma_{\lim} = \pi R_s^2 \tag{43}$$

In Fig. 4, we observe that the energy emission rate increases with an increase in the dilaton parameter $N$. The presence of the dilaton field increases the energy emission rate drastically in the Einstein Generalized Maxwell gravity. Also, the peak of the energy emission rate takes place in larger frequencies as $N$ increases.

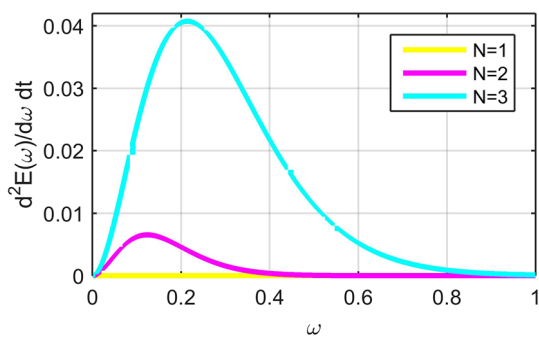

**Fig. 4** The energy emission rate against the frequency $\omega$. $m = 100$, $\Lambda = -1$, $p = q = 1$

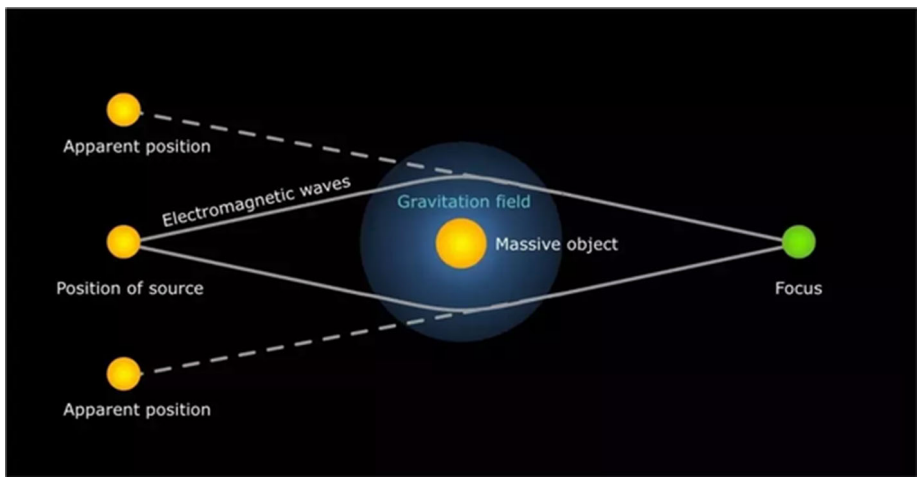

**Fig. 5** The schematic figure of gravitational lensing phenomenon

## 5 Gravitational lensing

Gravitational lensing is one of the first fruits of General Relativity [25]. This phenomenon was recognized in deflections of light by the sun (see Fig. 5). After that, this interesting event was captured by quasars, galaxy clusters, etc. Now, this phenomenon is ordinary in astronomical observations [26].

5.1 Deflection angle

A generic spherically symmetric metric has this line-element:

$$ds^2 = -A(x)dt^2 + B(x)dx^2 + C(x)d\Omega^2 \quad (44)$$

The photon sphere is a radius that photons can orbit the black hole unstably. This radius can be obtained as the largest root of the following equation

$$\frac{A'(x)}{A(x)} = \frac{C'(x)}{C(x)} \quad (45)$$

The deflection angle can then be calculated as a function of the closest approach for the photons that coming from infinity has the form [27]

$$\alpha(x_0) = I(x_0) - \pi \tag{46}$$

where

$$I(x_0) = 2 \int_{x_0}^{\infty} \frac{\sqrt{B(x)}dx}{\sqrt{C(x)}\sqrt{\frac{C(x)A(x_0)}{C(x_0)A(x)} - 1}} \tag{47}$$

It is found that the deflection angle grows when $x_0$ approaches $x_{ps}$, and there is a special point that the deflection angle reaches $2\pi$ which means the photon winds a complete loop around the black hole. To discuss the divergence, one could introduce a new variable $z$ [28]

$$z = \frac{A(x) - A(x_0)}{1 - A(x_0)} \tag{48}$$

the integral in Eq. (12) becomes

$$I_D(x_0) = \int_0^1 R(0, x_{ps}) f_0(z, x_0) dz \tag{49}$$

where the functions $R(0, x_{ps})$ and $f_0(z, x_0)$ can be defined through the following equations

$$R(z, x_0) = \frac{2\sqrt{A(x)B(x)}}{A'(x)C(x)}[1 - A(x_0)]\sqrt{C(x_0)} \tag{50}$$

$$f(z, x_0) = \frac{1}{\sqrt{A(x_0) - [(1 - A(x_0))z + A(x_0)]C(x_0)[C(x)]^{-1}}} \tag{51}$$

$$f_0(z, x_0) = \frac{1}{\sqrt{\phi(x_0)z + \gamma(x_0)z^2}} \tag{52}$$

where

$$\phi(x_0) = \frac{1 - A(x_0)}{A'(x_0)C(x_0)}[A(x_0)C'(x_0) - A'(x_0)C(x_0)] \tag{53}$$

and

$$\gamma(x_0) = \frac{[1 - A(x_0)]^2}{2[A'(x_0)]^3[C(x_0)]^2}\Big[2[A'(x_0)]^2 C(x_0)C'(x_0) - A(x_0)A''(x_0)C(x_0)C'(x_0)$$
$$+ A(x_0)A'(x_0)\big[C(x_0)C''(x_0) - 2[C'(x_0)]^2\big]\Big] \tag{54}$$

With these definitions, Eq. (14) can be separated into two parts as:

$$I(x_0) = I_D(x_0) + I_R(x_0) \tag{55}$$

These parts are the divergent part $I_D(x_0)$ and the regular part $I_R(x_0)$.
The regular part $I_R(x_0)$ is defined as

$$I_R(x_0) = \int_0^1 \big[R(z, x_0)f(z, x_0) - R(0, x_{ps})f(z, x_0)\big]dz \tag{56}$$

$$c_1 = \frac{R(0, x_{ps})}{2\sqrt{\gamma(x_{ps})}} \tag{57}$$

and

$$c_2 = -\pi + c_R + c_1 \ln \frac{2\gamma(x_{ps})}{A(x_{ps})} \tag{58}$$

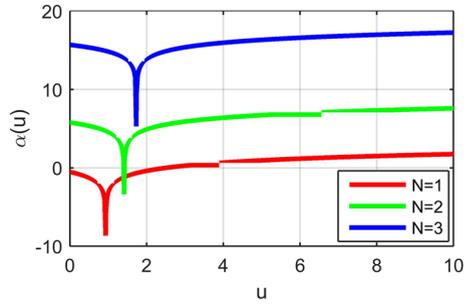

**Fig. 6** Plot showing the variation of deflection angle as a function of impact parameter $u$ for different values of $N$ when $m = 1$, $\Lambda = -1$, and $q = 0$

with

$$c_R = I_R(x_{ps}) \tag{59}$$

The deflection angle is plotted as a function of the impact parameter in Fig. 6 for different values of other parameters and it diverges as $u \to u_m$. Before this critical point, the deflection angle decreases with an increase in impact parameter and increases after this point as the impact parameter increases.

### 5.2 Positions and magnifications

The lens geometry setup is shown in the following figure (Fig. 7). Light emitted by the source is deflected by the lens and detected by an observer at an angle $\theta$. The background geometry is taken asymptotically flat. Also, both the source and the observer are placed in the flat space-time. The lens equation can be written as [29]:

$$\tan \beta = \tan \theta - \frac{D_{ds}}{D_s}[\tan \theta + \tan(\alpha - \theta)] \tag{60}$$

In the strong-field regime, when the source, lens, and observer are highly aligned, the lens equation takes this form:

$$\beta = \theta - \frac{d_{ds}}{d_s} \Delta \alpha_n \tag{61}$$

and the deflection angle reads:

$$\alpha(\theta) = -c_1 \ln\left(\frac{d_d \theta}{u_{ps}} - 1\right) + c_2 \tag{62}$$

by obtaining $\theta$ in terms of $\alpha$ and Taylor expansion around $\alpha = 2n\pi$, the angular position of the $n$-th image takes the following form:

$$\theta_n = \theta_n^0 - \zeta_n \Delta \alpha_n \tag{63}$$

where

$$\theta_n^0 = \frac{u_{ps}}{d_d}[1 + e^{(c_2 - 2n\pi)/c_1}], \quad \zeta_n = \frac{u_{ps}}{c_1 d_d} e^{(c_2 - 2n\pi)/c_1} \tag{64}$$

the angular positions of the images are

$$\theta_n = \theta_n^0 + \frac{\zeta_n d_s}{d_{ds}}(\beta - \theta_n^0) \tag{65}$$

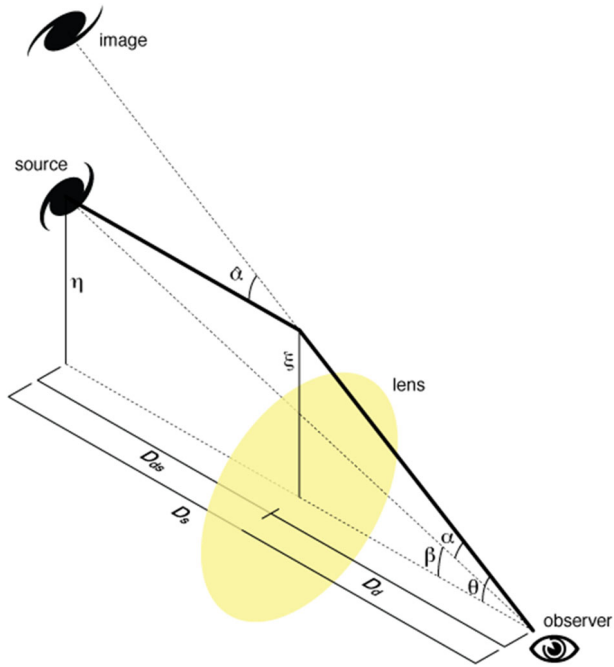

**Fig. 7** Gravitational lensing diagram

an infinite sequence of Einstein rings with angular radii

$$\theta_n^E = \left(1 - \frac{\zeta_n d_s}{d_{ds}}\right)\theta_n^0 \tag{66}$$

The magnification of the $n$-th relativistic image is given by (approximated to first order of $\zeta_n d_s/d_{ds}$)

$$\mu_n = \frac{1}{\beta}\frac{\theta_n^0 \zeta_n d_s}{d_{ds}} \tag{67}$$

5.3 Observables for the galactic black holes

We are going to define the recognizable observables [30]. The angular position of the outermost image is $\theta_1$ and the other images packed together at:

$$\theta_\infty = \frac{u_{ps}}{d_d} \tag{68}$$

then

$$s = \theta_1 - \theta_\infty \approx \theta_\infty e^{(c_2 - 2\pi)/c_1} \tag{69}$$

and

$$r = \frac{\mu_1}{\sum_{n=2}^{\infty}\mu_n} \approx e^{2\pi/c_1} \tag{70}$$

where $s$ is the separation between the outermost image and the others, and $r$ is the flux ratio between the outermost image and all the others.

Table 1 $M = 2.8 \times 10^6 M_\odot$, $D_d = 8.5$ kpc

|  | $N = 1$ | $N = 2$ | $N = 3$ |
|---|---|---|---|
| $r$ | 6.8218 | 6.8219 | 6.8220 |
| $\theta_\infty$ | 16.86 | 1275.77 | 242,357.11 |
| $c_1$ | 1 | 0.99999 | 0.99998 |
| $u_{ph}$ | 2.5980 | 196.4779 | 37,324.5682 |

The center of our galaxy is believed to host a black hole with mass $M = 2.8 \times 10^6 M_\odot$ and its distance from us is about $D_d = 8.5$ kpc [31]. In this work, we have modeled the center of our galaxy using a black hole metric in dilaton gravity inspired with nonlinear Maxwell invariant. In the following table, we have obtained some observables to our model which in the case of $N = 1$, the Schwarzschild model recovers.

As is clear, one could see the impact of dilaton field ($N > 1$) on observables in the above table.

## 6 Concluding remarks

In this paper, we have studied the shadow cast and strong field gravitational lensing for a new class of black hole solutions in dilation gravity where the dilaton field is coupled with a nonlinear Maxwell invariant. In the presence of dilaton field, the Hamilton–Jacobi equation (26) takes a general form as (28) wherein the case of $N = 1$, we reach the standard form of Hamilton–Jacobi equation. We have shown that the shadow size increases with an increase in the dilaton parameter $N$ regardless of $q = 0$ or $q \neq 0$. Also, the energy emission rate increases with an increase in this parameter. After that, we have calculated several strong field lensing parameters such as the angular position ($\theta_\infty$), the ratio between the flux of the first and other images ($r$), constant parameter ($c_1$). The impact of the $N$ parameter on the lensing parameters can be seen in Table 1.